\documentclass[twocolumn]{aastex631}          % referee version: for submission

\usepackage{CJK}
\usepackage{graphicx}

\begin{document}

\begin{CJK*}{UTF8}{gbsn}

\title{The Hierarchical Structure Of The Open Cluster NGC 752}

\correspondingauthor{Heng Yu}
\email{yuheng@bnu.edu.cn}

\author{Delong Jia(贾德龙)}
\affiliation{School of Physics and Astronomy, Beijing Normal University, Beijing, 100875, China}

\author[0000-0001-8051-1465]{Heng Yu(余恒)}
\affiliation{School of Physics and Astronomy, Beijing Normal University, Beijing, 100875, China}

\author[0000-0001-8611-2465]{Zhengyi Shao(邵正义)}
\affiliation{Shanghai Astronomical Observatory, Chinese Academy of Sciences, Shanghai, 200030, China}
\affiliation{Key Lab for Astrophysics, 100 Guilin Road, Shanghai, 200234, People's Republic of China}

\author[0000-0002-0880-3380]{Lu Li(李璐)}
\affiliation{Shanghai Astronomical Observatory, Chinese Academy of Sciences, Shanghai, 200030, China}

\begin{abstract}
The structure of open clusters provides key insights into their evolution and the dynamics of the Milky Way. Using Gaia DR3 data, we applied a hierarchical clustering algorithm to the open cluster NGC 752 based on the kinematical information and identified four substructures corresponding to different stages of disintegration. 
The cluster exhibits a pronounced signature of mass segregation. Its outer members show a clear expansion trend with a velocities of 0.25 $\rm{km~s^{-1}}$ in the plane of the sky. In addition, the system shows a projected rotational pattern with an angular velocity of approximately 0.03 $\rm{rad~Myr^{-1}}$.
We also identified a correlation between the escape times of disturbed members and the epochs at which the cluster crossed the Galactic disk, highlighting the role of Galactic tidal forces in accelerating cluster dissolution.
We conclude that hierarchical clustering based on projection bounding energy is effective for studying the internal structure of star clusters, but it has limitations when dealing with unconstrained structures such as tidal tails.\end{abstract}

\section{Introduction}           %% first-level sections will be auto-capitalized
\label{sect:intro}

Open clusters originate from the fragmentation and collapse of massive molecular clouds \cite{2003ARA&A..41...57L}, 
typically comprising hundreds of stars \citep{2008MNRAS.389..869E}. These stars are generally coeval, forming within a short timescale and sharing similar initial chemical compositions, which makes open clusters invaluable laboratories for studying stellar formation and evolution under same initial conditions. During evolution, clusters undergo processes such as mass segregation, dynamical relaxation, tidal disruption, and eventually dissolve into the Galactic field\citep{2019ARA&A..57..227K,2025ApJ...982L..43L}.

Gravitational interactions with the Galactic potential or nearby massive structures can lead to the formation of tidal tails, consisting of stars that gradually escape the cluster's gravitational binding.
These tails typically form leading and trailing components.

Tidal tails typically exhibit low densities, broader spatial distributions, and a wider range of proper motions. This sparse distribution makes tidal tails particularly susceptible to contamination from field stars, which further complicating the identification of tidal tail members\citep{2021A&A...645A..84M}. Early identification of tidal tails in open clusters primarily relied on stellar density distributions and color-magnitude diagrams. However, due to limited data precision at the time, these methods often faced significant challenges in distinguishing genuine tidal tail structures from field star contamination\citep[e.g.][]{2002A&A...389..871D,2008MNRAS.389..678B, 2015MNRAS.449.1811D, 2019AJ....157...12B}.

Gaia provide high-precision astrometric and kinematic data, including stellar positions, proper motions, and parallaxes, which has significantly enhanced our ability to detect and characterize these structures \citep{2016A&A...595A...2G,2018A&A...616A...1G,2023A&A...674A...1G}. 
By combining spatial and kinematic information, it is now possible to disentangle tidal tail members from the surrounding stellar population more effectively. 
For instance, \cite{2018A&A...618A..93C} used Gaia DR2 identified members and derived homogeneous parameters for 1229 open clusters with unprecedented precision, mapping their spatial and age distributions across the Galactic disk.
\cite{2022MNRAS.517.3525B} used Gaia EDR3 data to identify multiple known tidal tail cluster members. 
Also based on Gaia EDR3 data, \cite{2022A&A...659A..59T} systematically extended open cluster membership lists using the HDBSCAN algorithm, revealing that most clusters possess extended coronae and tidal tails, which provide new insights into cluster evolution and dissolution processes.
\cite{2025arXiv250417744X} selected a sample of open clusters with minimal line-of-sight distortions using Gaia DR3 data.

Grouping stars based on their characteristics can provide valuable insights into cluster evolution and the formation of tidal tails.
Existing member selection algorithms utilize multiple stellar attributes to identify cluster members, such as UPMASK\citep{2014A&A...561A..57K}, HDBSCAN\citep{2017JOSS....2..205M}, SHiP \citep[a FoF-based method,][]{2019ApJS..245...32L}. They primarily determine membership as a binary classification or assign a membership probability to each star. These methods, however, offer limited capability in distinguishing substructures within the cluster.
\cite{2021MNRAS.505.1607B} previously proposed using the tidal radius as a criterion to differentiate structural components within a cluster, introducing a new perspective for classifying cluster members.
To classify different groups of cluster members, \cite{YuShao-5} applied a hierarchical clustering algorithm for member selection. This method utilizes stellar positions and proper motions while revealing the hierarchical structure within the cluster. Such an approach facilitates the investigation of internal substructures and the relationship between the cluster and surrounding stars.
\cite{2024AJ....168...79J} applied this method to the young cluster NGC 6530, identified multiple substructures and studied their origin and evolution through kinematic information.

In order to test the potential of the method for studying the disintegration process of the open cluster,
we selected the old cluster NGC 752 known for its tidal tail as our target. This cluster has an age of approximately 1.4 Gyr, with a distance of about 430 pc from the Sun\citep{2023A&A...673A.114H}. Numerous studies have not only confirmed the presence of a tidal tail in this cluster but also carried out detailed investigations into its properties.
\cite{2021MNRAS.505.1607B} investigated the mass loss occurring in both the core and tidal regions of NGC 752 and estimated the current mass of the cluster to be $297 \pm 10$ M$_\odot$. Based on simulations, \cite{2022MNRAS.514.3579B} suggested that the tidal tails of NGC 752 may extend much farther than previously observed. \cite{2024A&A...686A.215L} further conducted a detailed analysis of the properties of blue dispersed stars within the tidal tails of NGC 752.

In this work, we apply the hierarchical clustering algorithm to NGC 752 to analyze its properties, and evaluate whether this approach is suitable for detecting tidal tails.
The structure of this paper is as follows: Section \ref{sec:method} introduces the hierarchical clustering algorithm used for member selection; Section \ref{sec:data} presents the data and the identified members of different substructures; Section \ref{sec:property} analyzes the physical properties of different substructures;
finally, Section \ref{sec:conclusion} provides a summary.

\section{Method} \label{sec:method}

In this section, we introduce the method used to select and group cluster members. This approach has been applied in previous studies to identify and distinguish members of the Perseus Double Cluster\citep{YuShao-5} and NGC 6530 \citep{2024AJ....168...79J},see \citet{2022A&C....4100662Y} for a more general review.

The method utilizes the projected binding energy between stars to perform clustering, arranging the stars into a binary tree (dendrogram). The hierarchical structure of the tree reflects the relative gravitational binding of stars within the cluster. By cutting the tree at appropriate levels, we can categorize members based on their gravitational binding states.

The projected binding energy between two stars is expressed as follows:

\begin{equation}
    E_{ij}=-G\frac{m_{i}m_{j}}{r\theta_{ij}}p+\frac{1}{2}\frac{m_{i}m_{j}}{m_{i}+m_{j}}\frac{{\triangle\mu_{x}}^2+{\triangle\mu_{y}}^2}{2}r^{2}
\end{equation}

\noindent In this equation, $\theta_{ij}$ represents the angular separation between the two stars, and \textit{r} denotes the distance from the star cluster to the observer. The parameters ${\triangle\mu_{x}}$ and ${\triangle\mu_{y}}$ correspond to the differences in proper motion along two perpendicular directions, where ${\triangle\mu_{x}=\triangle\mu_{RA}}$ and ${\triangle\mu_{y}=\triangle\mu_{DEC}}$ ($\mu_{RA}$ and $\mu_{DEC}$ are from Gaia). $m_{i}$ and $m_{j}$ show the mass of the stars. However, since mass estimation is affected by uncertainties in the cluster's distance, magnitude, and extinction, we adopt a uniform assumption of $m_{i}=m_{j}=1 M_{\odot}$ for all stars. Considering possible contributions from undetected members and the non-uniformity of velocity projection in all directions, a parameter $p$ is introduced to regulate the relative contributions of potential and kinetic energy within a reasonable range.

Then, we can construct the binary tree based on the projected binding energy between stars.
The main branch of the tree is identified by selecting the node with the largest number of leaves at each level. By traversing along this main branch and calculating the velocity dispersion associated with each node, we generate a velocity dispersion profile for the gravitational potential field. Once the branch enters the interior of the cluster, its velocity dispersion does not easily change with the decrease in members. A plateau will thus appear in the velocity dispersion profile. 
We refer to this plateau as the "$\sigma$ plateau", as in galaxy cluster studies\citep{1999MNRAS.309..610D, 2011MNRAS.412..800S, 2015ApJ...810...37Y}. 
If the object contains multiple hierarchical structures with different velocity dispersions, there will be multiple plateaus.

To characterize these plateaus, we fit the histogram of velocity dispersion using a Gaussian Mixture Model (GMM) and determine the optimal number of components according to the Bayesian Information Criterion (BIC). The GMM assigns weights to each Gaussian component representing different plateaus. We then compare the weights of the most dominant Gaussian component, across different values of the parameter $p$.
The clustering result corresponding to the value of $p$ that has the maximum weight is ultimately selected as the basis for a subsequent analysis.
Then, the plateaus identified by the GMM could be used to group members.

\section{Data and Analysis} \label{sec:data}

\subsection{Data Set}

To reduce the computational complexity, we performed an initial data selection. This selection was based on the cluster membership information provided by \cite{2023A&A...673A.114H}, with an expanded selection range to ensure broader coverage. We extract data from Gaia DR3 within a rectangular region centered at RA:29°, DEC:38°, with a field of 30° by 24°. 
We filter stars with a G-band magnitude brighter than 20 and restricted their proper motion to $6 < {\mu}_{RA} < 13$ and $-15 < {\mu}_{Dec} < -8$ (mas/yr). Furthermore, by restricting the parallax to the range of 1.5 to 3 mas, we ultimately obtained a sample of 19,359 stars.
Among these stars, only 3,690 have radial velocity measurements,  with parallax information and their mean parallax uncertainty is about 0.21 mas, corresponding to an uncertainty of roughly 50 pc at this distance, which is much larger than the size of the cluster. 
Therefore, in the subsequent analysis, we adopt only four dimensional kinematical data: RA, DEC and corresponding proper motions.

\subsection{Cluster Membership}
\label{ssub}

\begin{figure}
\label{fig:w-p}
\centering
\includegraphics[width=\linewidth]{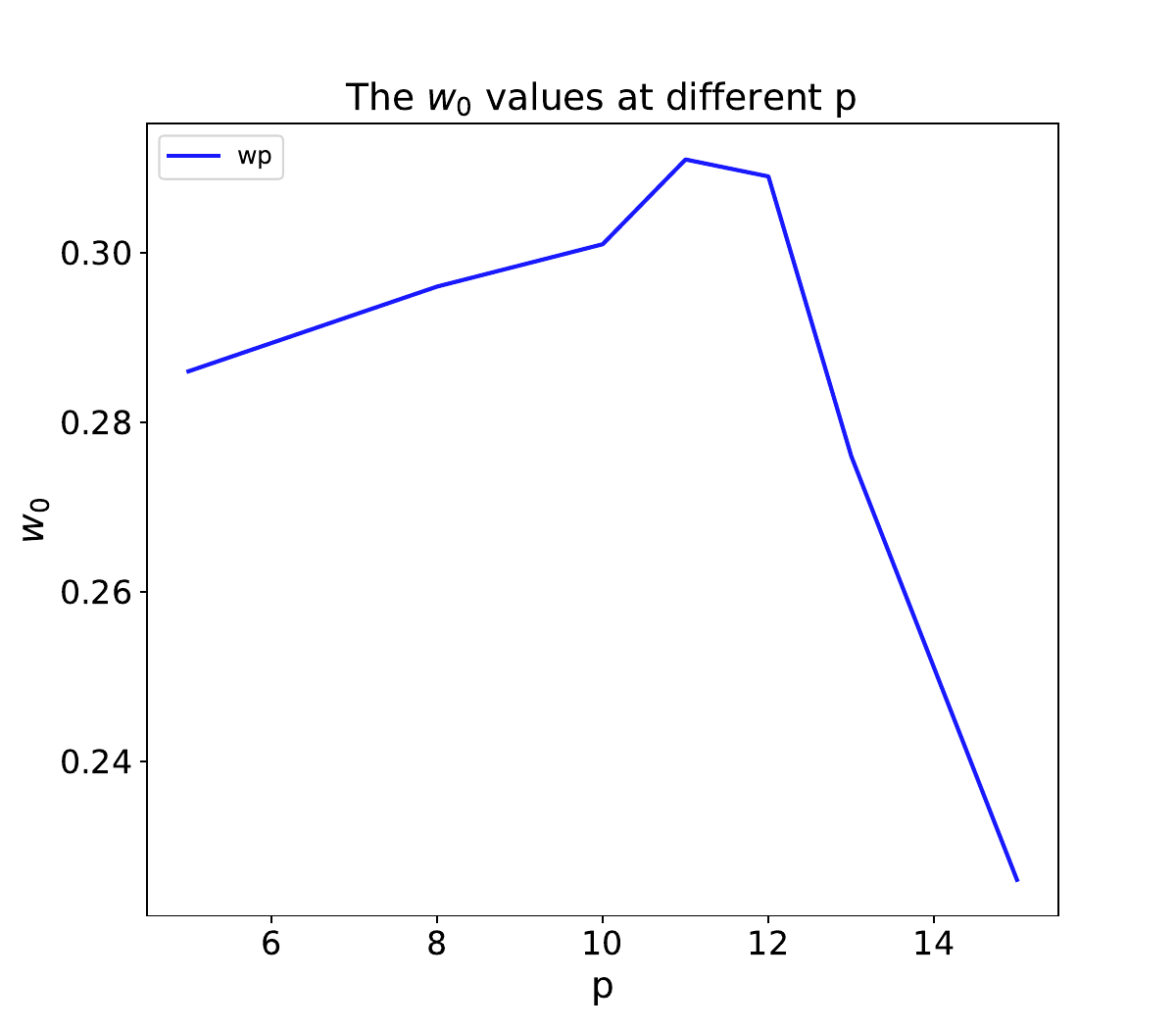}
\caption{The $w_0$-$p$ plot. $w_0$ represents the proportion of the largest plateau when the algorithm performs Gaussian fitting on different plateaus.}
\end{figure}

The first step of the hierarchical clustering algorithm is to determine the parameter $p$. We collect the weight values ($w_0$) of the $\sigma$-plateau with different $p$. 
The plot depicting the variation of $w_0$ with respect to the $p$ parameter is shown in Figure \ref{fig:w-p}. The figure reveals that the maximum $w_0$ value occurs when the $p$ parameter is set to 11 . Therefore, we set $p=11$ for further analysis.

\begin{figure}
\centering
\includegraphics[width=\linewidth]{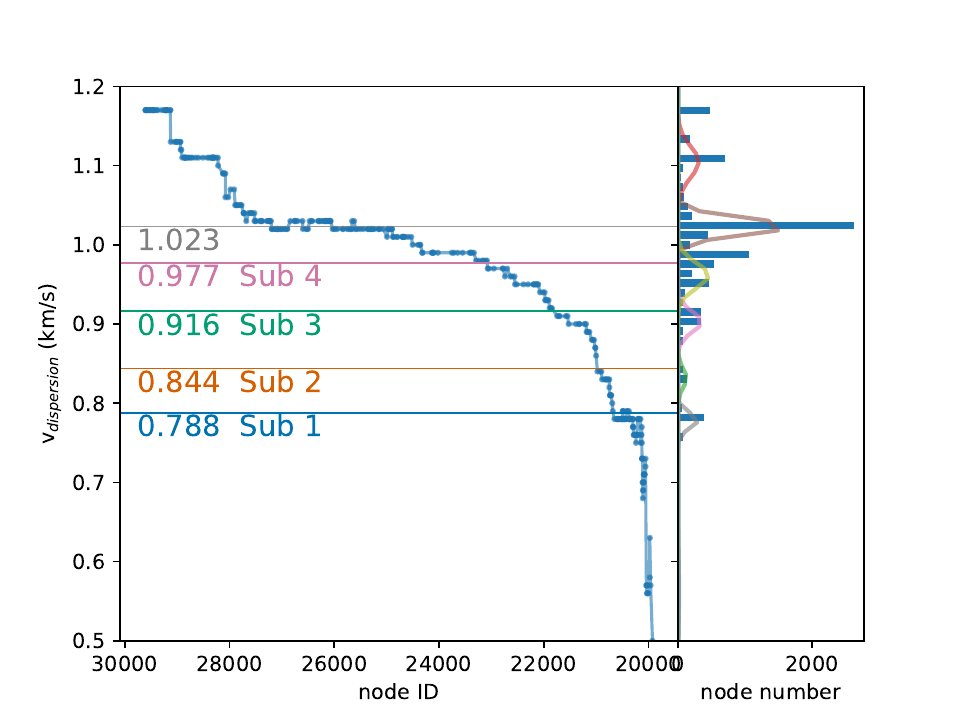}
\caption{\label{fig:vhst}{\it Left:} Velocity dispersion profile, with leaf nodes on the x-axis and velocity dispersion values on the y-axis. 
The gray line corresponds to largest velocity plateau 1.02 $\rm{km~s^{-1}}$.
Other four colored lines, from low to high, representing 0.788, 0.844, 0.916, and 0.977 $\rm{km~s^{-1}}$, represent the thresholds that separate different hierarchical substructures.
{\it Right:} Histogram of velocity dispersion, with solid colored lines representing multiple Gaussian components.}
\end{figure}

Then, the algorithm produced the velocity dispersion profile along the main branch of the binary tree within the field of view, as shown in Figure \ref{fig:vhst}. The largest velocity plateau is located at 1.023 $\rm{km~s^{-1}}$.
However, there are many foreground and background stars are located within the main plateau. To improve the purity of the members,  we cut the tree at a lower level.
We adopt the upper boundary (1$\sigma$ above the mean value, 0.977 $\rm{km~s^{-1}}$) of the first Gaussian component below the main plateau as a threshold. This yields a set of 388 member stars below this threshold.

\begin{figure*}
\centering
\includegraphics[width=\linewidth]{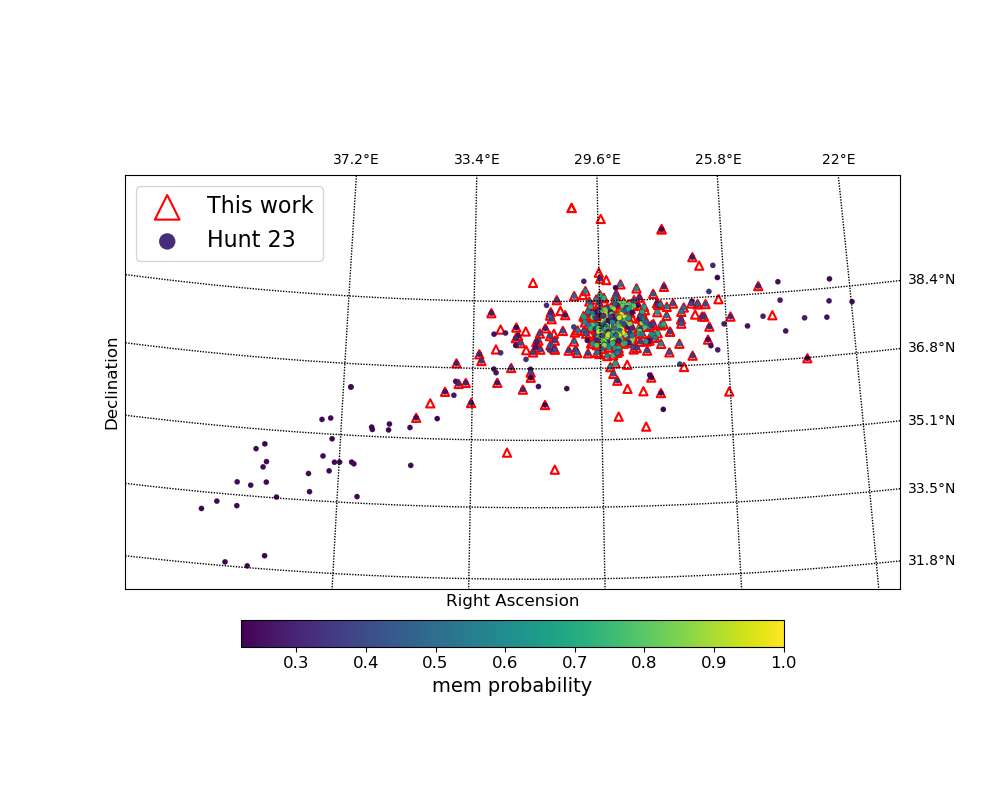}
\caption{\label{fig:space_h23} The member of the star cluster NGC 752 selected by H23 (solid dots) and this work (red blank triangles). The probabilities of H23 member stars are marked with colors.}
\end{figure*}

To verify our membership, we compare the identified members with the results of previous studies, as shown in Table \ref{tab:compare}.
For members identified by \cite{2020A&A...640A...1C}(hereafter C20), which using Gaia DR2 data and trained an artificial neural network using their photometry and parallaxes to select cluster member, we cross-match them based on their spatial coordinates using 1 arcsec as the criterion.
\citet{2023A&A...673A.114H} (hereafter H23) identified members of NGC 752 using Gaia DR3 data and HDBSCAN method, which provides membership probabilities for each star. We cross-match their catalog using Gaia source IDs and compared our selection with both their full member list and a subset with membership probability larger than 0.3.
The number of member stars identified in each study is denoted as $n_{\rm work}$, and the number of overlapping members with our selection is denoted as $n_{\rm match}$.

Table \ref{tab:compare} shows that the cluster members selected by our algorithm exhibit a high level of consistency with those identified by other methods(97\% of C20 members are included, and 79\% for H23). The members selected by \citet{2020A&A...640A...1C} have both lower member counts and smaller velocity dispersions, suggesting that they mainly concentrate in the inner region of the cluster. The high-probability members ($p_{mem}>0.3$) from H23 exhibit very similar physical properties to our members. However, when including members with lower membership probabilities, the overall velocity dispersion of the H23 sample increases to 1.50 $\rm{km~s^{-1}}$. 
As shown in Fig. \ref{fig:space_h23}, these less reliable members are mainly located in the outskirts of the cluster.

The HDBSCAN algorithm employed by H23 is an unsupervised clustering technique based on the similarity of stellar parameters. It is designed to identify groups of stars with comparable physical properties within large datasets. Consequently, the method can also detect stars that are physically similar to cluster members but located at larger distances, such as potential escapees that are no longer gravitationally bound to the cluster. By contrast, our method primarily targets bound members and is less effective at tracing escaped stars.

\begin{table}[ht]
  \centering
  \caption{Comparison of Membership Selection Results for NGC 752 in Different Studies}
  \label{tab:compare}
  \begin{tabular}{|l|c|c|c|c|c|}
    \hline
    Work & $n_{\rm mem}$ & $n_{\rm match}$ & $v_{\rm dis}$  & $\mu_{\rm RA}$  & $\mu_{\rm Dec}$  \\
    &             &             &($\rm{km \cdot s^{-1}}$) & \multicolumn{2}{c|}{($\rm{mas \cdot yr^{-1}}$)}\\
    \hline
    C20 & 223 & 216 & 0.78 & 9.81 & -11.71 \\
    H23 & 436 & 343 & 1.50 & 9.71 & -11.92 \\
    H23{\footnotesize ($>$0.3)} & 317 & 300 & 0.96 & 9.79 & -11.84 \\
    This work & 388 & - & 0.97 & 9.77 & -11.85 \\
    \hline
  \end{tabular}
\end{table}

\subsection{The Kinematical Structures}

\begin{figure}
\centering
\includegraphics[width=\linewidth]{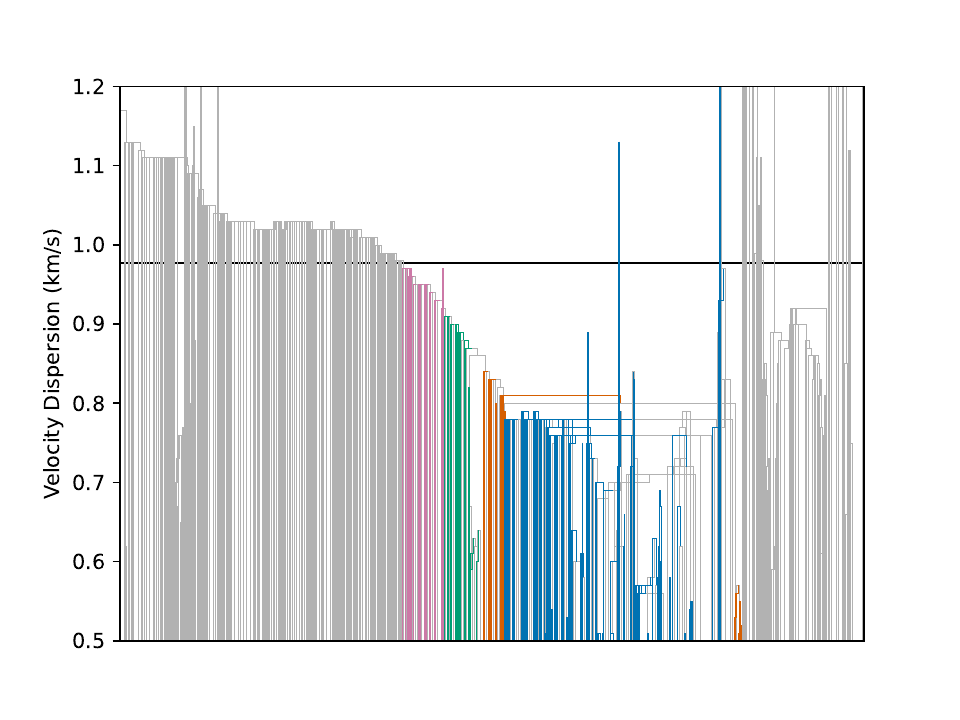}
\caption{\label{fig:dtree} The dendrogram of NGC 752, with the y-axis representing velocity dispersion. Blue indicates the Sub 1, orange Sub 2, green Sub 3, and pink Sub 4.}
\end{figure}

To further investigate the kinematical structure of the cluster, we focused on the velocity plateau below 0.977 $\rm{km~s^{-1}}$ in the velocity dispersion diagram. Similar to previous steps, cutoff thresholds were set at +1$\sigma$ above the centers of the corresponding Gaussian components (0.788, 0.844, and 0.916 $\rm{km~s^{-1}}$), resulting in four substructures, named Sub 1, Sub 2, Sub 3 and Sub 4. The distribution of these substructures within the binary tree is shown in Figure \ref{fig:dtree}. And some basic physical properties of each substructure are listed in Table \ref{tab:subs}.

\begin{table}[htpb]
    \centering
    \caption{Basic properties of substructures}
    \label{tab:subs}
    \begin{tabular}{|c|c|c|cc|}
        \hline
        & n  & $v_{dis}$& ${\mu}_{RA}$ & ${\mu}_{DEC}$  \\
        & & ($\rm{km \cdot s^{-1}})$ & \multicolumn{2}{|c|}{($\rm{mas \cdot yr^{-1}}$)}  \\
        \hline
        Sub 1 & 262 & 0.78& $9.78\pm0.26$ & $-11.84\pm0.28$   \\
        Sub 2 & 33 & 1.12& $10.03\pm0.38$ & $-11.87\pm0.39$    \\
        Sub 3 & 45 & 1.27& $9.71\pm0.47$ & $-11.84\pm0.40$    \\
        Sub 4 & 48 & 1.28& $9.65\pm0.48$ & $-11.80\pm0.40$    \\
        \hline
    \end{tabular}
\end{table}

\begin{figure*}
\centering
\includegraphics[width=\linewidth]{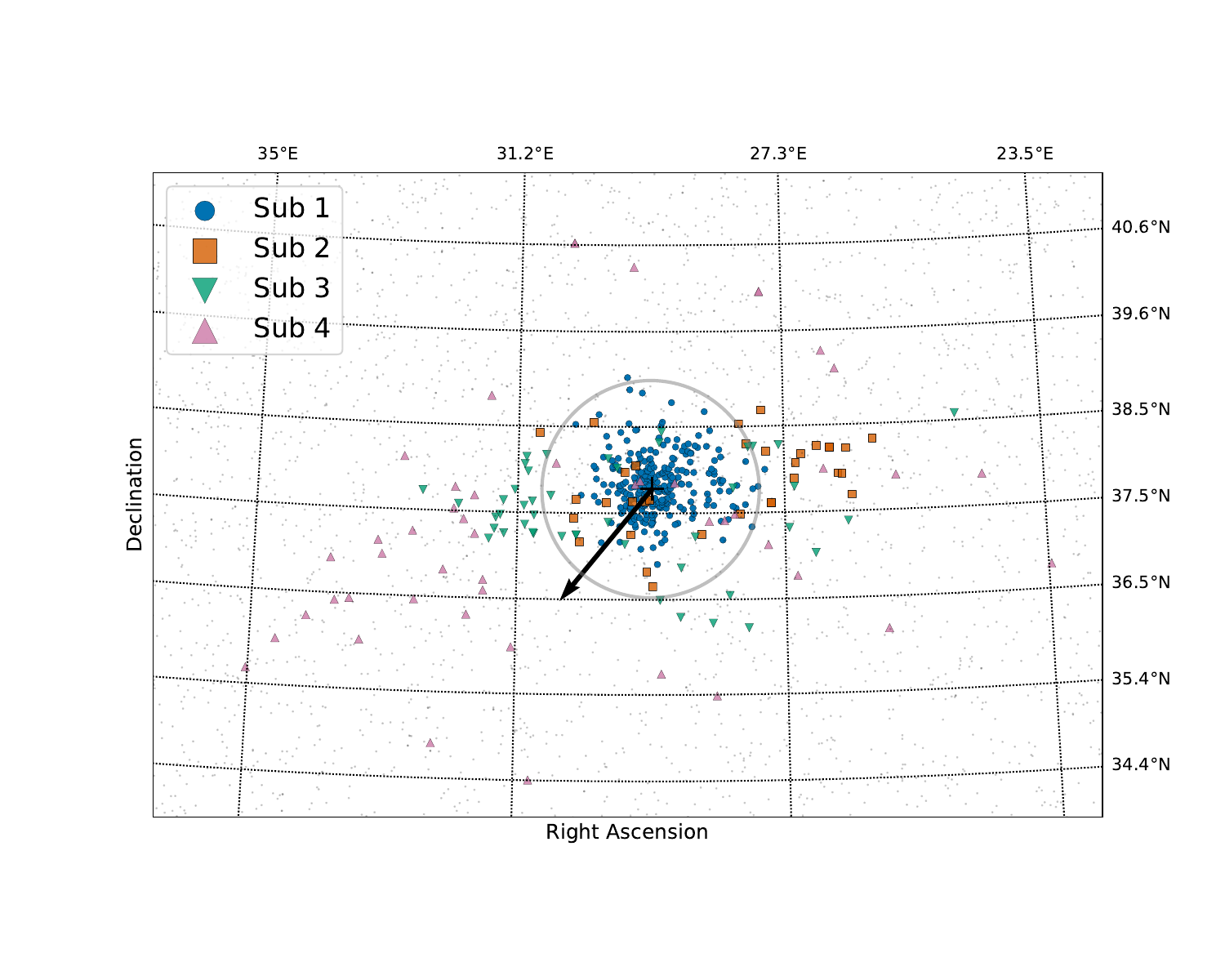}
\caption{\label{fig:space} The spatial distribution of four substructures of NGC 752. Blue circles represent the Sub 1, orange squares represent the Sub 2, green inverted triangles represent the Sub 3, and pink triangles represent the Sub 4. The center of the cluster is indicated by a black plus sign, and the tidal radius(9.5 pc) is indicated by a gray circle.}
\end{figure*}

The most intuitive property is their spatial distribution on the celestial sphere, as shown in Figure \ref{fig:space}. The figure displays the distribution of the four substructures in celestial coordinates and includes reference circles centered on the cluster center (RA: 29.26°, Dec: 37.78°) with a tidal radius of 9.5pc, as given by \citet{2021MNRAS.505.1607B}. 

It is evident that the Sub 1 is located in the central region of the cluster, presenting a concentrated high-density core. 
The Sub 2 is primarily distributed in the west and also includes some stars spatially close to the core. Sub 3 lies mainly to the east, with its members extending away from the cluster. In contrast, Sub 4 appears more diffuse, consists of weakly bound members, and contains a tidal tail extending to the southeast.

\begin{figure}
\centering
\includegraphics[width=\linewidth]{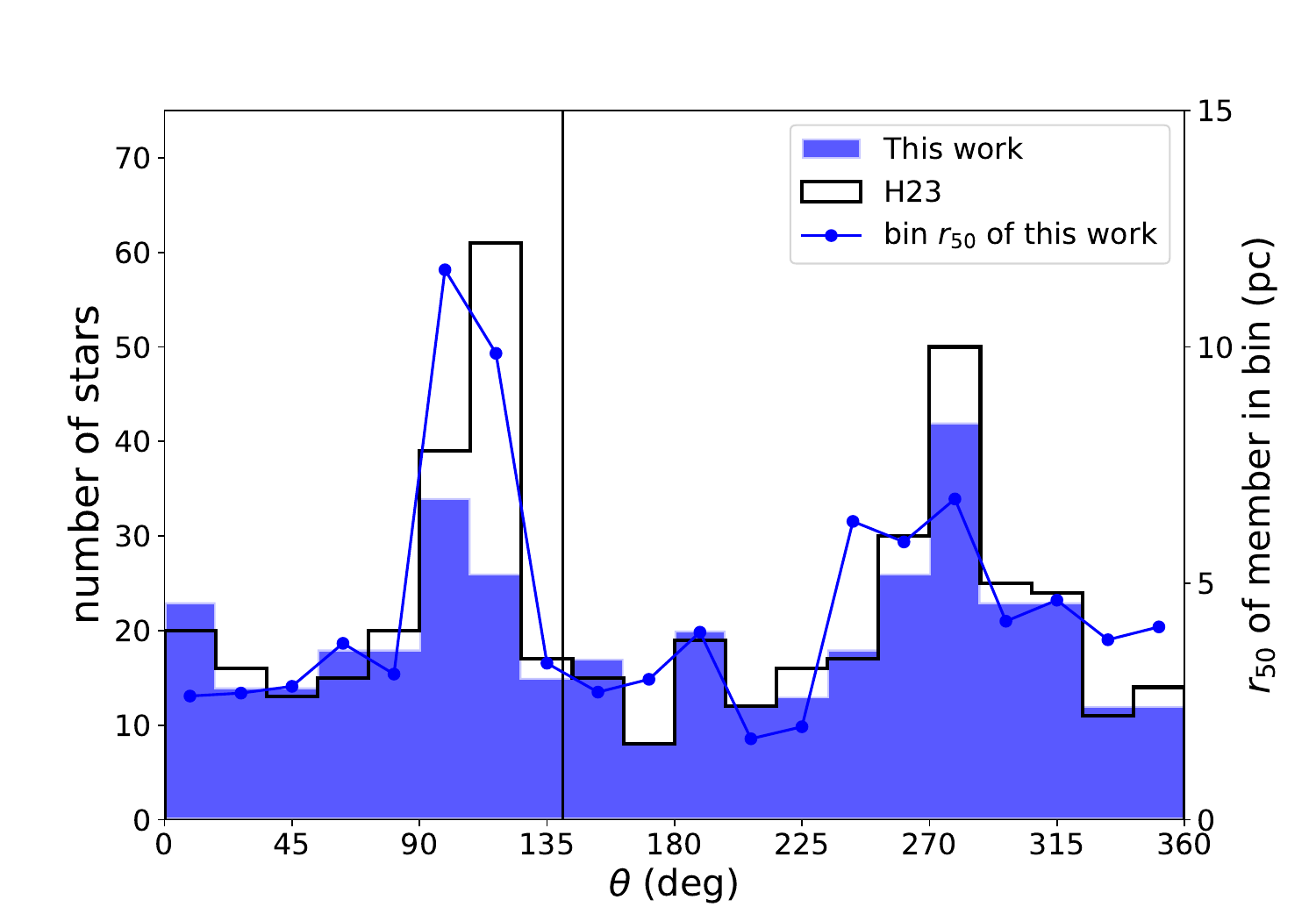}
\caption{\label{fig:angle}The histogram of the number of cluster member stars as a function of position angle, along with the median distance of members from the cluster center ($r_{50}$) within each angle bin. The blue bins are from member stars of our method, while gray bins from members of \citet{2023A&A...673A.114H}. Position angles $\theta$ are measured counterclockwise from north through east relative to the north.
The black vertical line represents the direction of motion of the cluster as a whole.}
\end{figure}

To quantify the spatial distribution of the cluster's tidal tails, we plotted a histogram (Figure \ref{fig:angle}) showing the proportion of cluster members as a function of position angle, as well as the median distance of members from the cluster center (defined as $r_{50}$) within each angle bin. 
We define north as 0°, with angles increasing counterclockwise.
The results reveal a significant overdensity of member stars along the tidal tail direction (100° and 280°). The star number histogram indicates that
the east tail is close to the direction of the cluster's motion, contains more members, and extends farther. This is consistent with observations in other clusters that the leading tidal tail contains more stars and is longer than the trailing tail\citep{2024Kroupa}. 
This result also suggests that our method misses some members that have escaped.

\begin{figure}[htp]
\centering
\includegraphics[width=\linewidth]{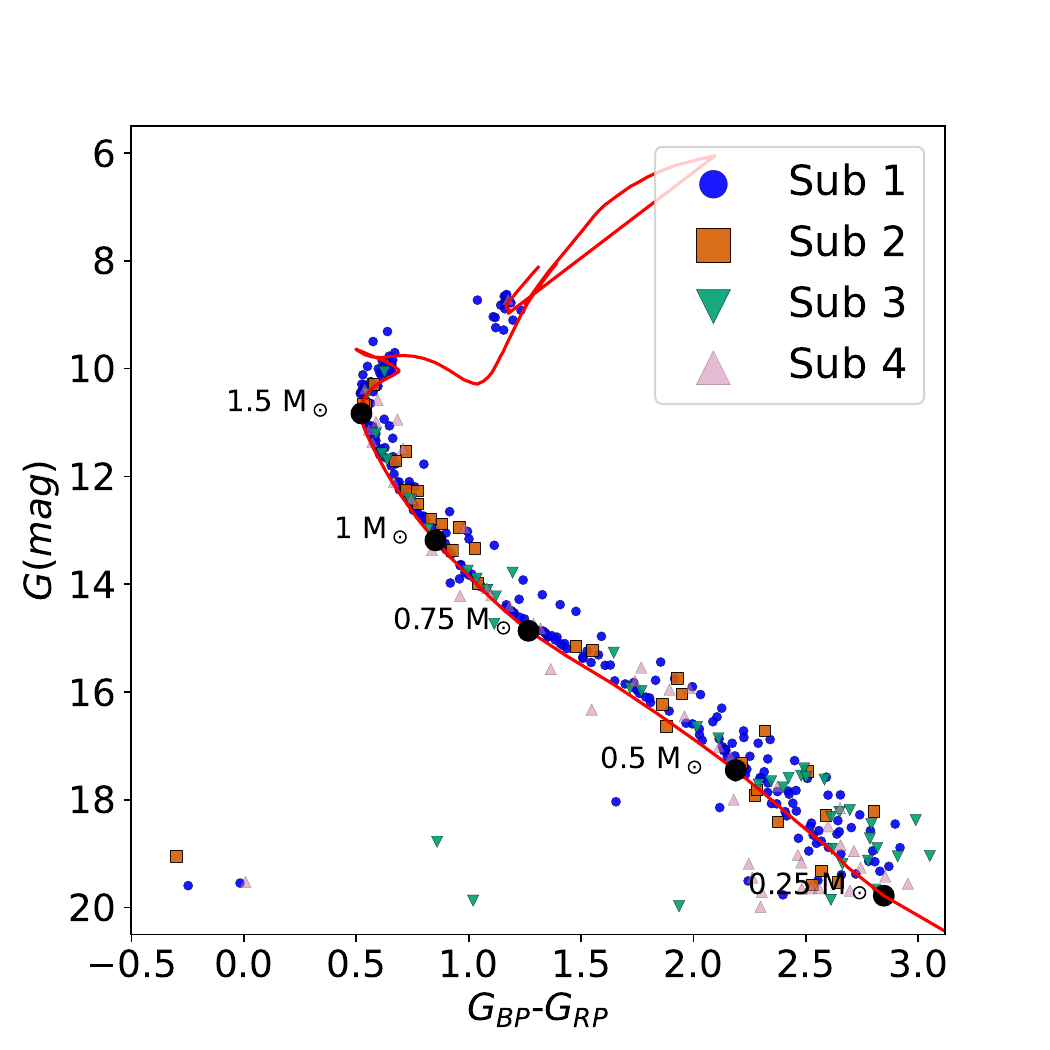}
\caption{\label{fig:hr}
Color-magnitude diagrams of all members. The color code is the same as Fig. \ref{fig:space}. The red line represents the isochrone of 1.74 Gyr.}
\end{figure}

The relationship between stellar color and magnitude reflects stellar mass and evolutionary stage. For a coeval population originating from the same molecular cloud, the stars should approximately follow the same isochrone left envelope in the color–magnitude diagram (CMD). Therefore, we plotted the CMDs of our member stars in Figure \ref{fig:hr}. They exhibit a clear left envelope, consistent with the typical characteristics of an old and dynamically evolved cluster.

We adopted the PARSEC theoretical isochrones \citep[version 1.2S][]{2012MNRAS.427..127B} along with the Gaia DR3 photometric system \citep[][G, GBP, and GRP bands]{2021A&A...649A...3R}. We use the following input parameters from the MiMO catalog \citep{2025AJ....170..288L}: [Fe/H] = –0.058, distance modulus = 8.166, and visual extinction $A_V$ = 0.149. The final fit yields an estimated age of approximately 1.74 Gyr for NGC 752. Furthermore, by interpolating the mass for each member star from the isochrone using its G-band magnitude, we derived the masses of the substructures are 245.2, 26.3, 27.9, and 33.1 M$_\odot$, respectively, with a total cluster mass of 332.5 M$_\odot$. This mass estimation should be considered as a lower limit, as our interpolation method does not account for unresolved binaries. Assuming a 30\% binary fraction with a uniform mass ratio distribution, we find our mass is likely underestimated by approximately 16\%.

Overall, NGC 752 has a relatively dense core(Sub 1), the outer members are being gradually stripped away(Sub 2 \& Sub 3), and there is a distinctly elongated tidal tail to the southeast(Sub 4). Based on the kinematic information of their members, we can make a more in-depth study of the evolutionary trend of this cluster.

\section{Physical Property}\label{sec:property}

\subsection{Mass Segregation}

During the long-term dynamical evolution of the cluster, gravitational interactions among stars, dynamical friction, and stellar mass differences lead to a process known as mass segregation. As a result, massive stars tend to concentrate toward the cluster center, while lower-mass stars migrate outward and may eventually be stripped away, forming tidal tails.

\begin{figure}
\centering
\includegraphics[width=\linewidth]{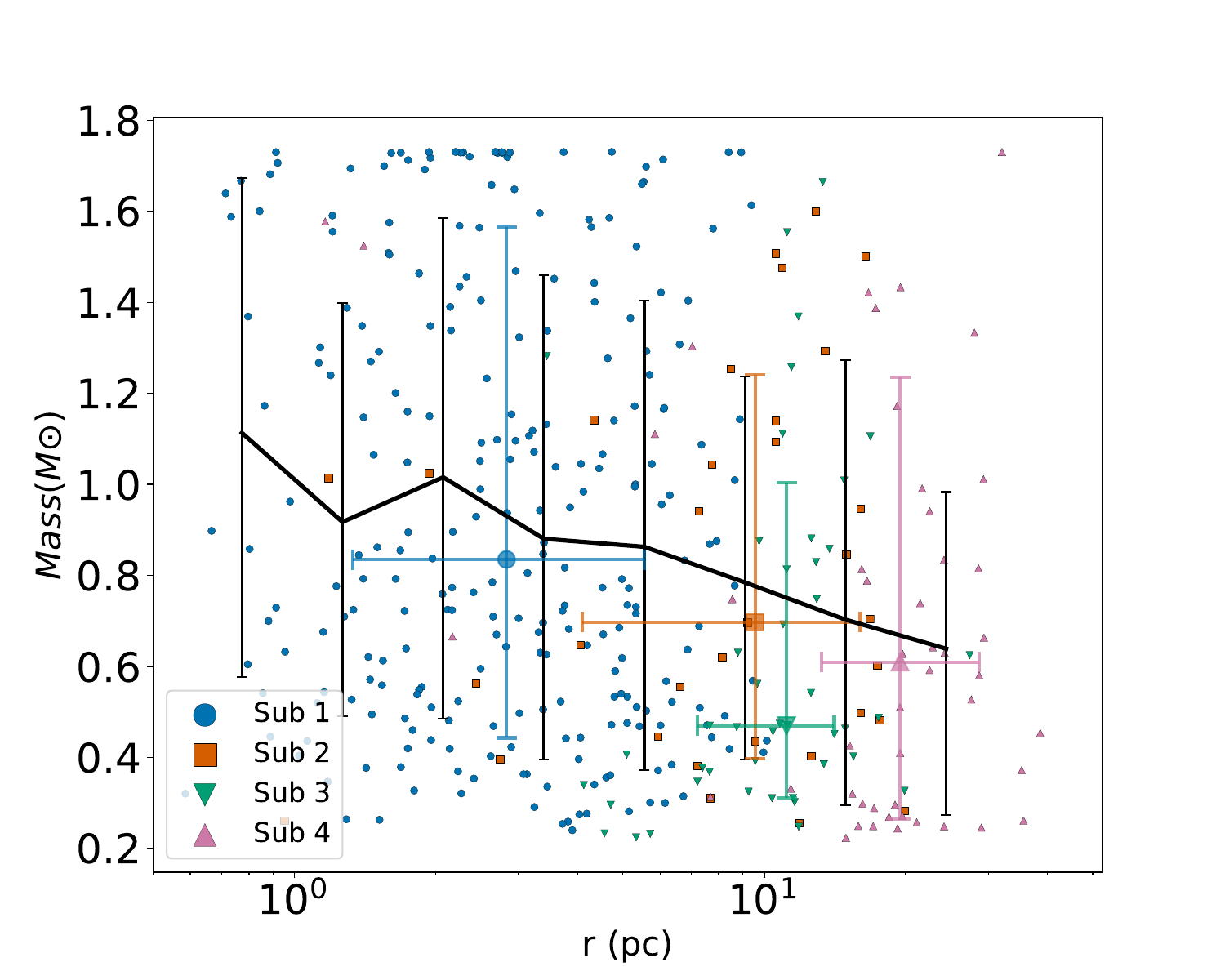}
\caption{\label{fig:segre}The masses of all cluster members with their distances from the center of the cluster. The average values of each substructure are marked with errorbars. The black error bars indicate the mean and standard deviation of masses in distance bins.
}
\end{figure}

To investigate whether mass segregation has occurred in NGC 752, we analyzed the relationship between stellar mass and distance from the cluster center, as shown in Figure \ref{fig:segre}. In addition, we plotted the stellar masses of each substructure against their distances from the cluster center. To illustrate the trend of mass as a function of distance, we binned the data from this work by distance and calculated the mean and standard deviation of the stellar masses within each bin. These results are shown as error bars. The results clearly indicate that lower-mass stars are more likely to be distributed in the outer regions of the cluster, which is consistent with the typical phenomenon of mass segregation. This suggests that NGC 752 may have undergone a period of internal dynamical evolution.

\subsection{Radial Expansion}

\begin{figure}
\centering
\includegraphics[width=\linewidth]{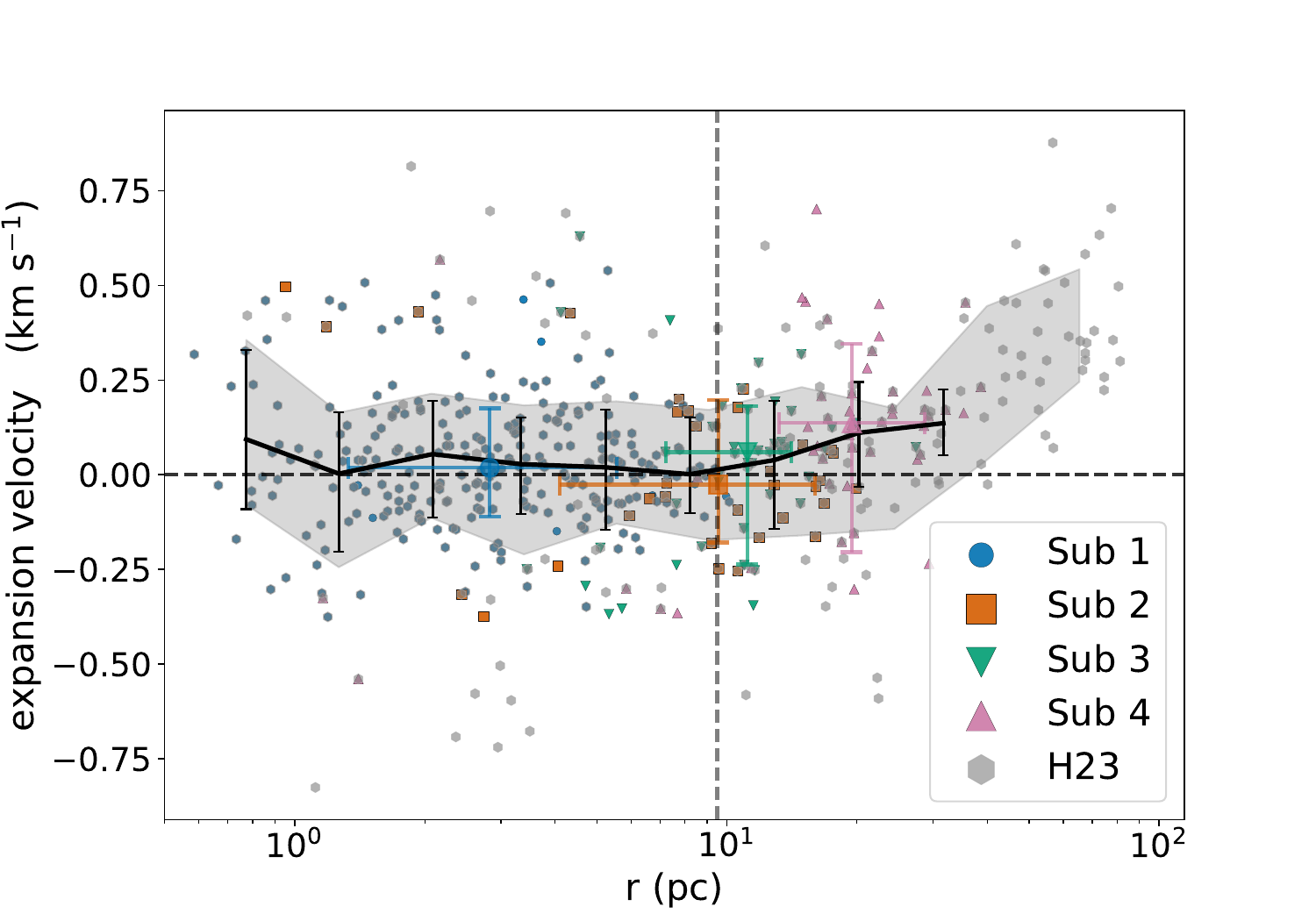}
\caption{\label{fig:vr}Expansion velocities of members and their distances from the cluster center. The black dashed line indicates the mean value of expansion velocity. The mean expansion velocities and errors of different substructures are marked with colorful errorbars. The gray shaded region shows the expansion trend of H23 members.
}
\end{figure}

To investigate the possible expansion of NGC 752, we projected each star's proper motion along the radial direction relative to the cluster center. 
The trend of expansion velocities of member stars with respect to the projected radius ($r$) is shown in Figure~\ref{fig:vr}.
The trend represented by the shaded area is from the data of H23 members.
Stars within the tidal radius show no significant trend of expansion or contraction.
This indicates that the central region remains gravitationally bound. In contrast, beyond the tidal radius, a clear expansion pattern emerges -- radial velocities increase with distance from the center. This trend is also observed in more distant (beyond 25 arcmin) members identified by H23, supporting the interpretation that outer members undergo expansion likely driven by long-term external perturbations, consistent with the formation of the tidal tail.

\subsection{Rotation}

\begin{figure}
\centering
\includegraphics[width=\linewidth]{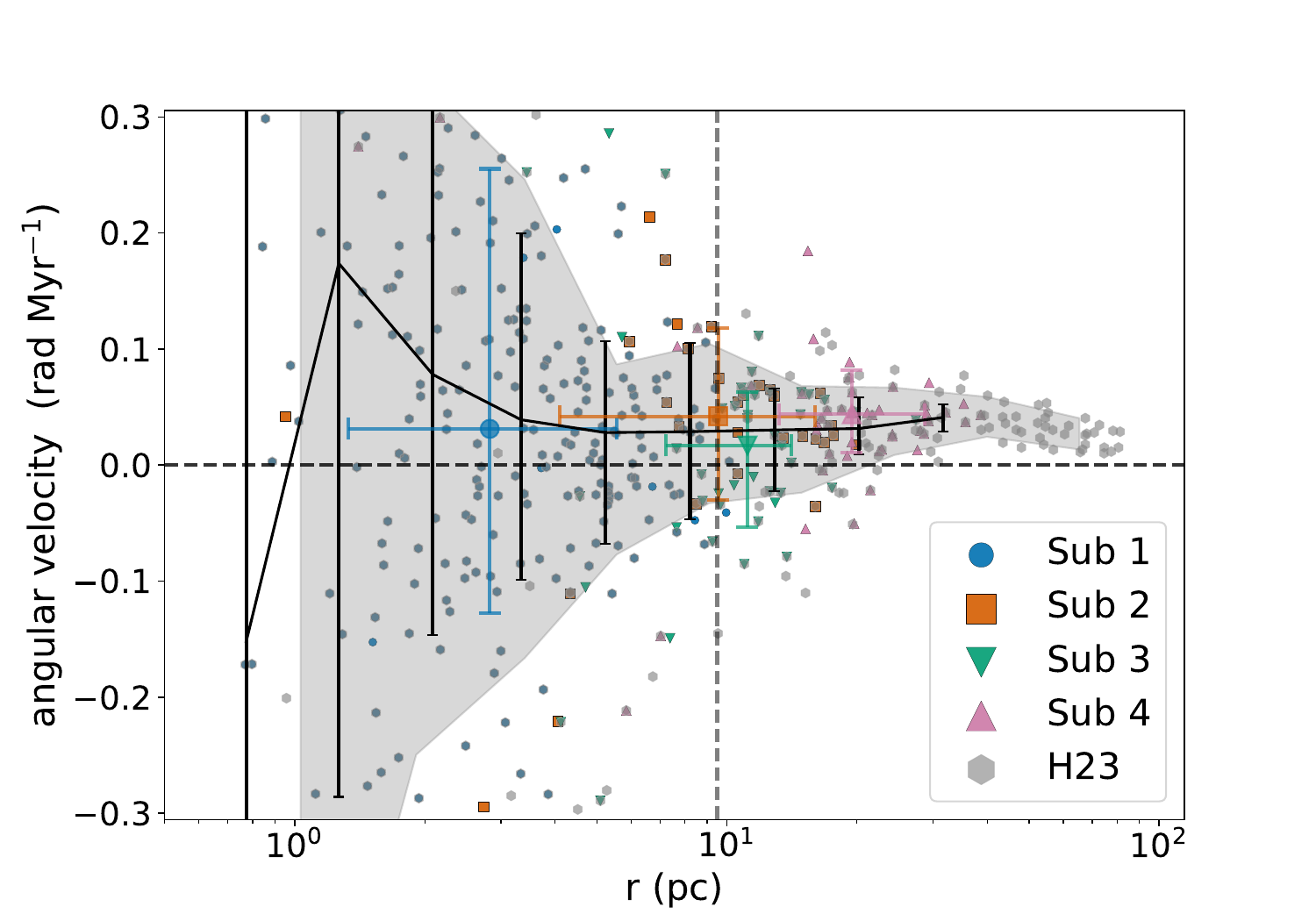}
\caption{\label{fig:va}The angular velocities of members and their distances from the cluster center.  The mean angular velocities and errors of different substructures are marked with colorful errorbars. The gray shaded region shows the rotation trend
of H23 members.
}
\end{figure}

We then analyzed the rotational behavior of stars on the sky plane by decomposing their tangential velocities and calculating the angular velocity (defined as tangential velocity divided by the projected radius), taking the clockwise motion as positive. Figure \ref{fig:va} shows that stars near the cluster center exhibit a large dispersion in angular velocity, indicating no coherent rotational pattern in the inner regions. However, stars located in the outer parts and within the tidal tails display a more consistent rotational motion, with angular velocities clustering around 0.029 $\rm{rad~Myr^{-1}}$. This suggests that the tidal tails of NGC 752 preserve a residual rotational signature. In contrast, the outermost stars from the H23 sample show a slight decline in angular velocity with increasing distance, implying that weakly bound stars gradually lose coherent rotation as they drift away from the cluster.

\begin{figure}
\centering
\includegraphics[width=\linewidth]{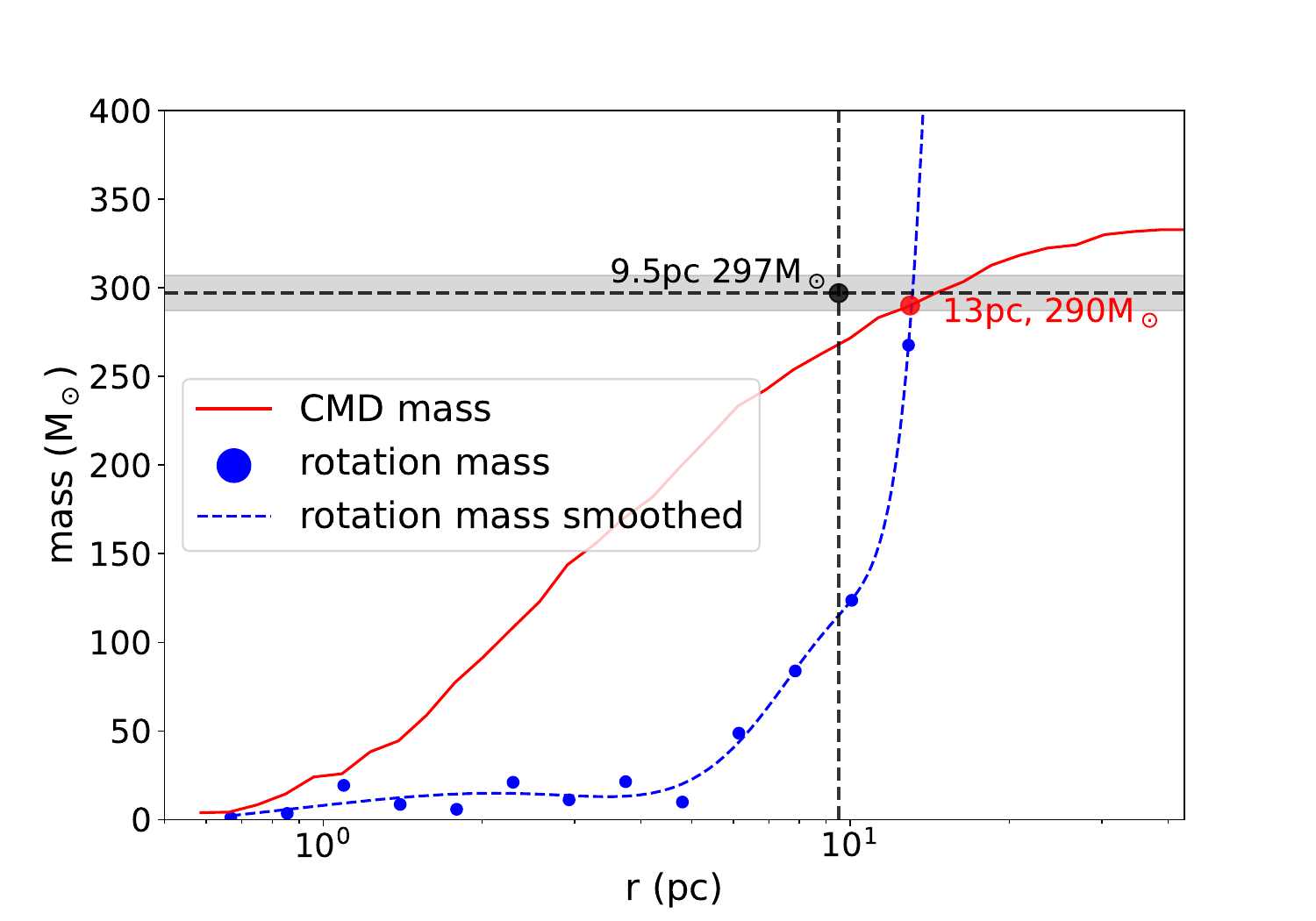}
\caption{\label{fig:mass}Mass profile from two approaches. The red line represents the total mass of member stars within different radii. Blue points indicate the mass required to sustain rotation at different radii. The blue line shows the polynomial fitted mass curve based on these points.
The black dashed line marks the reference tidal radius of 9.5 pc and a total mass of $297 \pm 10$ M$_\odot$ as given by \cite{2021MNRAS.505.1607B}.
}
\end{figure}

For a gravitationally bound system, the relationship between mass and rotational velocity can be derived using Newtonian mechanics. The total mass enclosed within a radius $R$ is related to the rotational velocity at radius $R$ by the following equation:

\begin{equation}
M(R) = \frac{v^{2} R}{G}=\frac{\omega^{2} R^{3}}{G}
\label{eq:mass}
\end{equation}

where $M(R)$ is the total mass within radius $R$, $\omega$ is the angular velocity of rotation, and $G$ is the gravitational constant.
This relationship allows us to estimate the mass required at each radius for the cluster to sustain rotation. We then compare this mass profile with masses derived from the color–magnitude diagram (CMD) analysis (see Figure \ref{fig:mass}). 
In the figure, the mass required for rotational support is fitted using a polynomial function, with the optimal polynomial degree determined by the BIC.
When the required mass to sustain rotation exceeds the CMD-inferred mass, the cluster can no longer remain gravitationally bound at that radius, and members located beyond this radius will gradually escape. 
The two curves intersect at about 13 pc, corresponding to a mass of 290 $M_\odot$. This result is consistent with results reported by \cite{2021MNRAS.505.1607B}: 9.5 pc and 297 $M_\odot$.
Since the rotational component in the line-of-sight direction is still unknown, our intersection position can only be taken as an upper limit on the tidal radius.

\subsection{Process of Disruption}

\begin{figure}
\centering
\includegraphics[width=\linewidth]{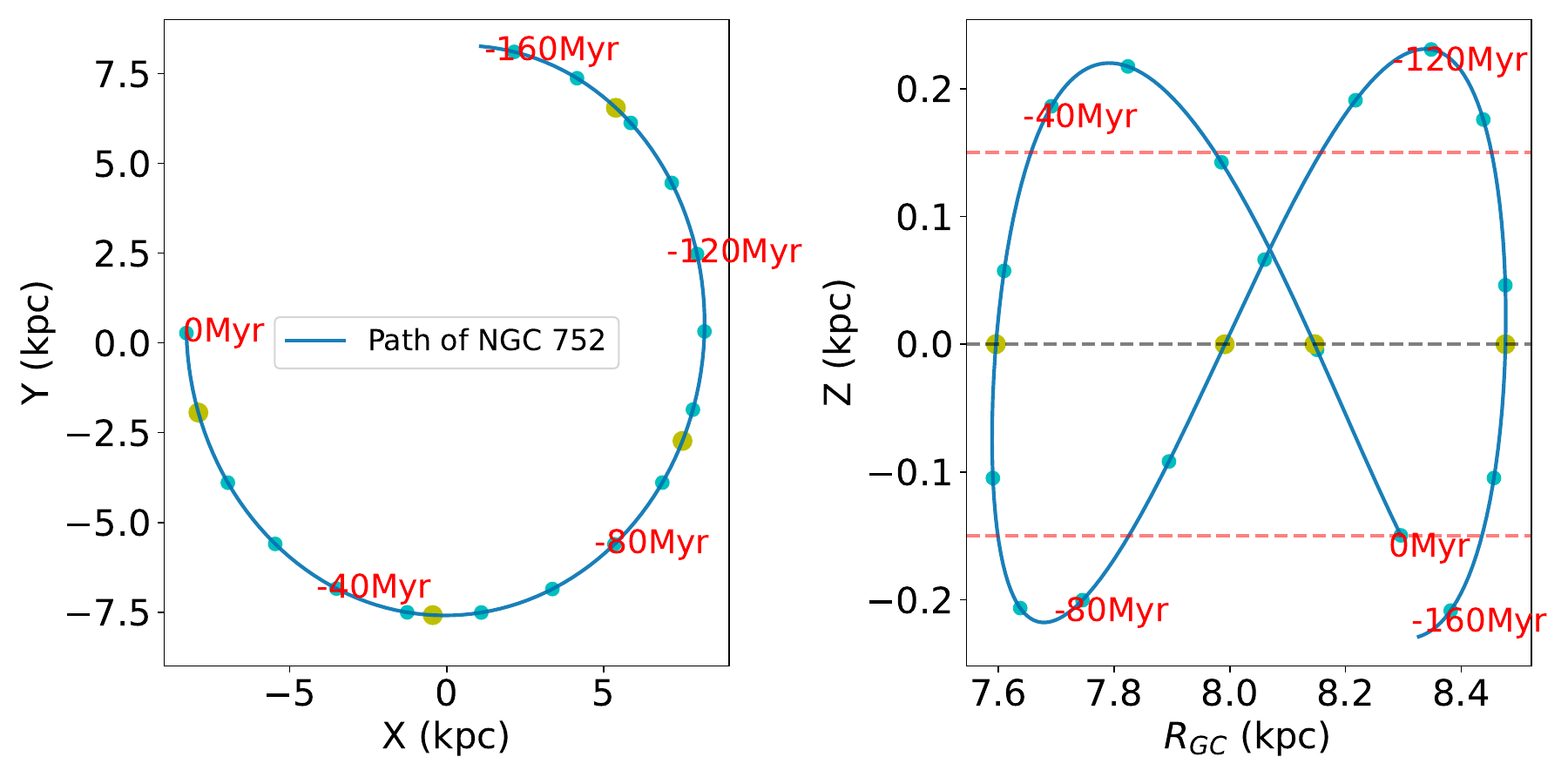}
\caption{\label{fig:path}Cluster orbit in the Milky Way. {\it Left:}  orbital trajectory in the XY plane. {\it Right:}  trajectory in the RZ plane, where R denotes the Galactocentric distance. Cyan dots indicate the position of the cluster at intervals of 10 Myr, and yellow dots mark the cluster's location when crossing the Galactic midplane (Z = 0). 
}
\end{figure}

\begin{figure}
\centering
\includegraphics[width=\linewidth]{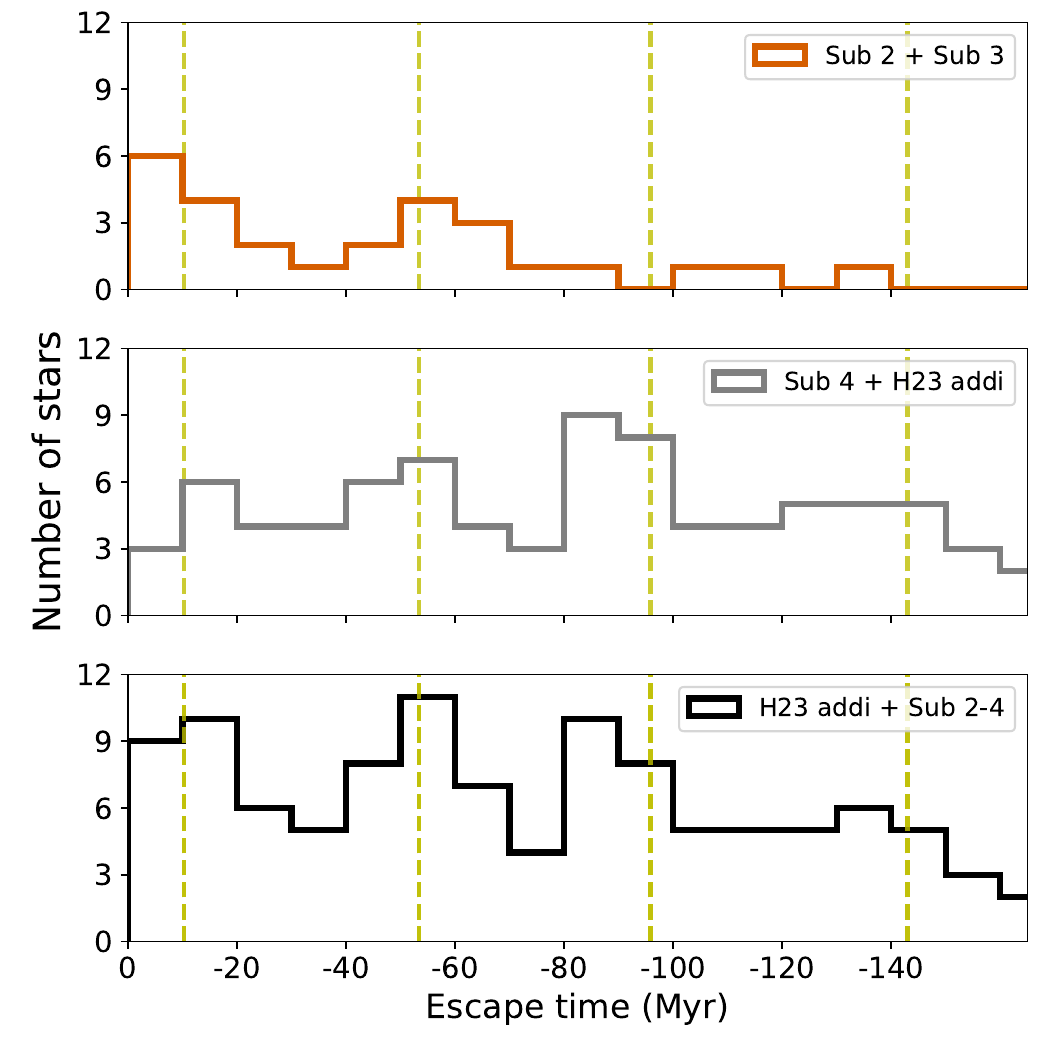}
\caption{\label{fig:back_time}The histogram of escape time for member stars from different structures. {\it top:} members of Sub 2 and Sub 3; {\it: middle:} Sub4 and H23 additional (H23 addi) members; {\it bottom:} all member stars except Sub 1. Yellow vertical dashed lines indicate the times at which the cluster crossed the Galactic midplane (Z = 0).
}
\end{figure}

To explore the origin of the tidal tail of NGC 752, we reconstructed the cluster's orbital path within the Milky Way using the \texttt{galpy} package with the MWPotential2014 model\citep{2015ApJS..216...29B}. 
Figure~\ref{fig:path} shows the cluster's trajectory through the Galaxy in the last 160 Myr, with cyan points marking positions at 10 Myr intervals and yellow points indicating the cluster's crossings of the Galactic midplane (Z = 0). 
There are four crossings occurring at 10 Myr, 53 Myr, 96 Myr, and 143 Myr ago. 
On the other hand, we select member stars located outside the tidal radius in substructures Sub 2–4 and those from the H23 catalog (comprising 93 additional members not included in our catalog, most of which are low-probability members, abbreviated as H23 addi). Assuming their expansion velocities have remained approximately constant in the recent past, we define the time of their most recent crossing of the tidal radius as the escape time. We group the members from Sub 2 and Sub 3, which are located at a distance from the cluster core, into one set, and those farther members from Sub 4 and H23 addi into another set. The histogram of escape times for the member stars in these groups is shown in Fig. \ref{fig:back_time}.

It shows that most stars in Sub 2–3 escaped from the cluster within the past 80 Myr. The peaks in the escape time histogram for these stars coincide with the cluster’s two most recent crossings of the Galactic disk.
For the members of Sub 4 and H23 addi, the peaks in their escape time histogram coincide with the cluster’s disk crossings at approximately 53 Myr and 96 Myr ago. For the earlier crossing that occurred 142 Myr ago, we did not find a corresponding excess in the number of escaping stars.
Our results indicate that the cluster’s disk-crossing events play a crucial role in perturbing its internal structure and stripping away member stars, as tidal forces are strongest near the Galactic plane. Moreover, earlier perturbation events are difficult to trace due to the subsequent disruption and dispersion of cluster members.

\section{Conclusions}\label{sec:conclusion}

In this study, we investigated the kinematic structure of the open cluster NGC 752 using Gaia DR3 data. By applying a hierarchical clustering algorithm, we identified four substructures at different levels of gravitational binding. Sub 1 corresponds to the cluster core, Sub 2 and Sub 3 represent stars in the process of detaching from the cluster on its western and eastern sides, respectively, and Sub 4 corresponds to a relatively bound component of the tidal tail. 
These members are generally consistent with the catalog of previous studies.

The cluster exhibits a pronounced signature of mass segregation. Its outer members show a clear expansion trend with a velocity of 0.25 $\rm{km~s^{-1}}$ in the plane of the sky. The cluster exhibits projected rotation with an angular velocity of 0.029 $\rm rad~Myr^{-1}$. Moreover, we identified a correlation between the escape times of disturbed members and the epochs at which the cluster crossed the Galactic disk, highlighting the role of Galactic tidal forces in accelerating cluster dissolution.

Beyond this specific case, our analysis demonstrates that hierarchical clustering methods are effective for disentangling the internal substructure of open clusters but are less suited to tracing escaped members. Complementary approaches—such as chemical tagging or detailed orbital modeling—will be necessary to follow their subsequent evolution.

\begin{acknowledgements}
This work has been supported by the National Key Research and Development Program of China (No. 2023YFC2206704)  and the China Manned Space Program with grant No. CMS-CSST-2025-A04. Zhengyi Shao acknowledges the National Natural Science Foundation of China (NSFC) under grants 12273091, U2031139, and the partly supports of the National Key R\&D Program of China No. 2019YFA0405501,  the science research grants from the China Manned Space Project with No. CMS-CSST-2021-A08 and the Science and Technology Commission of Shanghai Municipality (Grant No. 22dz1202400).
Li, L. thanks the support of NSFC No. 12303026 and the Young Data Scientist Project of the National Astronomical Data Center.

This work has made use of data from the European Space Agency (ESA) mission Gaia ( \href{https://www. cosmos.esa.int/gaia}{https://www. cosmos.esa.int/gaia} ), processed by the Gaia Data Processing and Analysis Consortium (DPAC, \href{https://www.cosmos.esa. int/web/gaia/dpac/consortium}{https://www.cosmos.esa. int/web/gaia/dpac/consortium} ). Funding for the DPAC has been provided by national institutions, in particular the institutions participating in the Gaia Multilateral Agreement. This research has made use of NASA's Astrophysics Data System Bibliographic Services. 

\software{astropy \citep{2013A&A...558A..33A,2018AJ....156..123A}, Scikit-learn\citep{2011JMLR...12.2825P}, galpy\citep{2015ApJS..216...29B}}
\end{acknowledgements}

\bibliographystyle{aasjournal}
\bibliography{bibtex}

\end{CJK*}
\end{document}